\documentclass[journal,twoside,web]{ieeecolor}
\usepackage{generic}
\usepackage{cite}
\usepackage{amsmath,amssymb,amsfonts}
\usepackage{algorithmic}
\usepackage{graphicx}
\usepackage{textcomp}

\usepackage[utf8]{inputenc}
\usepackage{amssymb,amsmath}
\usepackage{enumerate}
\usepackage{multirow}
\usepackage{subfigure}
\usepackage{microtype}
\usepackage{listings,multicol}
\usepackage{layouts}
\usepackage{url}
\def\BibTeX{{\rm B\kern-.05em{\sc i\kern-.025em b}\kern-.08em
    T\kern-.1667em\lower.7ex\hbox{E}\kern-.125emX}}
\usepackage{color}
\definecolor{orange}{rgb}{1, 0.35, 0}

\begin{document}
\renewcommand{\journalname}{IEEE JOURNAL OF BIOMEDICAL AND HEALTH INFORMATICS}
\newcommand{\thename}{OpenCLIPER}

\title{\thename: an OpenCL-based C++ Framework for Overhead-Reduced Medical Image Processing and Reconstruction on Heterogeneous Devices}
\author{Federico Simmross-Wattenberg, Manuel Rodríguez-Cayetano, Javier Royuela-del-Val, Elena Martín-González, Elisa Moya-Sáez, Marcos Martín-Fernández, and Carlos Alberola-López
	\thanks{This work has been submitted to the IEEE for possible publication. Copyright may be transferred without notice, after which this version may no longer be accessible.}
    \thanks{This work is supported by MINECO under grants TEC-2014-57428 and TEC2017-82408-R, as well as by Junta de Castilla y Le\'on under grant VA069U16.}
    \thanks{Federico Simmross-Wattenberg, Manuel Rodríguez-Cayetano,Elena Martín-González, Elisa Moya-Sáez, Marcos Martín-Fernández, and Carlos Alberola-López are with the Laboratorio de Procesado de Imagen (Image Processing Laboratory) at Universidad de Valladolid, 47011 Valladolid, Spain (e-mail: \{fedesim,manrod,emargon,emoysae,marcma,carlos\}@lpi.tel.uva.es). }
    \thanks{Javier Royuela-del-Val. is with Health Time Group, Córdoba, Spain (e-mail: j.royuela.v@htime.org).}
}

\maketitle

\begin{abstract}
Medical image processing is often limited by the computational cost of the involved algorithms. Whereas dedicated computing devices (GPUs in particular) exist and do provide significant efficiency boosts, they have an extra cost of use in terms of housekeeping tasks (device selection and initialization, data streaming, synchronization with the CPU and others), which may hinder developers from using them. This paper describes an OpenCL-based framework that is capable of handling dedicated computing devices seamlessly and that allows the developer to concentrate on image processing tasks.

The framework handles automatically device discovery and initialization, data transfers to and from the device and the file system and kernel loading and compiling. Data structures need to be defined only once independently of the computing device; code is unique, consequently, for every device, including the host CPU. Pinned memory/buffer mapping is used to achieve maximum performance in data transfers.

Code fragments included in the paper show how the computing device is almost  immediately and effortlessly available to the users algorithms, so they can focus on productive work. Code required for device selection and initialization, data loading and streaming and kernel compilation is minimal and systematic. Algorithms can be thought of as mathematical operators (called processes), with input, output and parameters, and they may be chained one after another easily and efficiently. Also for efficiency, processes can have their initialization work split from their core workload, so process chains and loops do not incur in performance penalties. Algorithm code is independent of the device type targeted.
\end{abstract}


\section{Introduction}
\label{sec:introduction}
In the last decades, Medicine has benefited enormously from medical imaging techniques. Starting with radiography and echography, and more recently X-ray tomography, nuclear medicine and magnetic resonance imaging, these modalities are now of widespread use in most hospitals all over the world. Since the advent of these techniques in clinical practice, huge effort has been carried out by scientists worldwide to make them faster, easier to use for the practitioner, more accurate and, overall, safer and more convenient for the patient. However, this comes at the price of higher computational costs to the extent that processing power is often the bottleneck.

Graphics Processing Units (GPUs), when used as general-purpose computing devices, may overcome these computational needs, as several software packages (see section \ref{sec:related_work} for a number of examples) have shown. Nevertheless, software development over GPU devices is not straightforward: parallelization itself poses a significant paradigm change from traditional (single-threaded) programming but, remarkably as well, GPUs are autonomous devices and, as such, work asynchronously from the CPU. Hence, communication and synchronization issues arise likewise, hindering the use of GPUs to develop new medical imaging algorithms, especially in the earliest research stages, in spite of their unquestionable value.

In addition, the sake of simplicity often causes developers to focus on only one device type (e.g. CPU vs. GPU), which, in turn, tends to result in augmented effort if multiple device types need to be supported afterwards. In this regard, it should be stressed that interesting processing devices beyond CPUs and GPUs are available, namely DSPs, FPGAs, or many-core systems such as Intel Phi. OpenCL comes up here as a natural answer to the problem of supporting multiple device types but, when compared to its direct competitor in the GPU field, nVidia CUDA, it tends to fall behind in terms of simplicity and maturity.

The aforementioned scenario, combined with the need to set up a complex working environment which is inherent to asynchronous programming, give rise to more particular, although closely related, issues: device selection and initialization, data streaming to/from the computing device, synchronization and communication between threads, memory management and others; those isssues bring forth an additional overhead that prevents researchers and practitioners from focusing on the algorithm development itself (the so-called \textit{kernels}).

This paper presents a C++ programming framework which intends to solve the forenamed issues while being:

\begin{itemize}
    \item \textit{Device agnostic}: being OpenCL-based, no knowledge about the final computing device type is needed at development time.
    \item \textit{Straightforward to use}: device discovery and initialization, data transfers to and from the device and the file system, kernel loading and compiling, etc. are handled automatically. Therefore, the developer focuses on kernel development while housekeeping chores are kept to a minimum.
    \item \textit{Coherent in code writing and data manipulation}: data structures need to be defined only once, independently of the final computing device types. Accordingly, host and device code is unique for all device types.
    \item \textit{Easy to integrate in existing projects}: device-dependent traits are confined in a class specifically designed for this purpose, while already existing classes for data processing and storage are sufficiently generic so as to support manipulation of many ordinary data types ---such as real and complex N-dimensional arrays, among others---, but also easy to override and to derive from, if needed.
\end{itemize}

The rest of this paper is organized as follows: section~\ref{sec:related_work} reviews the main existing approaches to the problem of using dedicated devices for medical image processing, with an special focus on image reconstruction; section~\ref{sec:architecture} describes the architecture of \thename; section~\ref{sec:case_study} shows a case study where \thename\ is used to reconstruct an MRI image from fully sampled $K$-space; we also include performance comparison of \thename with two other well-known packages as well as a comparison with OpenMP; section~\ref{sec:conclusion} concludes the paper.

\section{Related Work}
\label{sec:related_work}
The use of dedicated devices for medical image processing has drawn notable attention in  recent years, especially with the evolution of GPUs towards general-purpose computing devices (GPGPU). This section briefly describes the most remarkable works in this direction we are aware of.

The Berkeley Advanced Reconstruction Toolbox (BART) \cite{bart} is a programming library as well as a collection of command line tools to carry out MR  image reconstruction \cite{sedona16}. In the paper, the authors refer to this toolbox as a ``framework for iterative image reconstruction'', so the terms library and framework are used interchangeably albeit differences exist between both. This toolbox has a twofold orientation, namely, it serves for rapid algorithm prototyping and testing  and it also facilitates the integration of these algorithms into the data acquisition and reconstruction pipeline. The library also provides support for parallel computation using multiple CPUs or, for some of the available algorithms, GPUs. CUDA is, however, the only supported API for the latter option, so dedicated processing is limited to GPU-type devices from nVidia.

Another interesting and very popular framework is ``Gadgetron''\cite{gadgetron}; this framework  allows the user to construct streaming data processing pipelines, where data is processsed by a series of modules (the gadgets). The pipeline is configured by means of an XML description file. Gadgetron is based on a client-server architecture, in which the latter is responsible for performing the actual reconstruction computations while the former is in charge of sending raw data to the server and receiving reconstructed images from it, either for visualization or storage. Client-server communication works over TCP/IP, using an \textit{ad-hoc} protocol which typically conveys the input or output data and the XML description file. This XML file contains a reader and a writer definition as well as a list of gadgets which conform the processing pipeline. Gadgets are specified by their name, i.e. the library in which they are located and the actual class that implements it; properties can also be defined for parameter setup within each gadget. Gadgetron uses Linux container technology~\cite{containers} to provide its network services, and is distributed through the Docker platform~\cite{docker}.
Since it is conceived as a network service with a client-server approach, Gadgetron deployment should be carefully planned in advance. However, Gadgetron allows the practitioner to build gadgets as standalone applications as well, for cases where no client/server approach or pipeline specification is needed. Parallel computing is supported in CPU as well as in GPU \cite{TMISorensen2009}, again, via CUDA.

Other well-known software packages for MR acquisition and reconstruction are Yarra~\cite{Yarra}, \verb$CS_MoCo_LAB$~\cite{MoCoLab}, Codeare~\cite{codeare}, MIRT~\cite{MIRT} and GPUNufft~\cite{gpuNUFFT}. However, we do not mean to be exhaustive since our aim is not to overcome these platforms in terms of performance, but we intend to solve the issues referred to in the introduction while complying with relevant requirements for frameworks (also itemized in that section) and maintaining performance comparable to other state of the art frameworks.

Apart from these initiatives, general-purpose software packages are available, such as Matlab or Python, which have been adapted to GPU computing by adding a CUDA interface (by means of the Parallel Computing Toolbox in Matlab, and of the PyCUDA package for Python). For the latter, PyOpenCL~\cite{pyopencl} wraps the OpenCL API so it can be invoked from Python.

With some similarities with the former libraries, GPI \cite{gpi} can also be mentioned. It is a Python-based graphical programming interface used for image reconstruction. The interface is typically used to create workflows for data manipulation and algorithm testing; built-in functionality can be extended by means of simple code interfaces. Recently, some functionality of BART seems to be integrated within a GPI node library.

Overall, Gadgetron seems a nice tool with a similar motivation as BART. i.e., rapid algorithm prototyping and testing, although Gadgetron is more easily seen as a framework while BART may be thought of as a library or a collection of tools. In this sense, \thename\ has been designed as a framework, with development of new algorithms in mind. On the other hand, computation on dedicated devices does not seem to be at the core of either Gadgetron or BART. Alas, being CUDA-based, algorithm development differs substantially depending on the final computing device (which can only be the CPU or a nVidia GPU). Hence, both BART and Gadgetron include non-interchangeable versions of the same algorithm written for CPU and GPU. This limitation lies deeply within the CUDA fundamentals and is, therefore, difficult to overcome. Typically, it manifests itself as the need of two different data structures to hold the same data: one for \textit{host} data (for CPU computing) and another for \textit{device} data (for GPU computing). Gadgetron, for instance, has \texttt{hoNDArray} for host and \texttt{cuNDArray} for GPU, as does Matlab with its GPU-specific \texttt{GPUArray}. BART, on the other hand, relies on conditional compiling to select the desired variant at compile-time. The use of OpenCL overcomes this limitation, as the same source code and data structures serve their purpose independently of the final computing device (which is chosen at run-time), but this particular piece of freedom comes at the price of complexity, due to the broader range of circumstances which must be taken into account: multiple vendors with their own extensions and limitations, multiple device types, multiple device instances, intercommunication between devices and the host, choosing the optimum device for a particular task, and so on. \thename\ takes care of most of them. 

OpenCL is a sound, versatile and efficient platform to support computing on dedicated devices, and is an industry standard supported by the Khronos Group~\cite{khronos}. However, it is more complex and less mature than CUDA, two facts that hinder its widespread usage in the medical image computing community, despite its advantages. \thename\ has been designed to fill in this gap by managing the complexity referred to above and offering developers direct interaction with device kernels, host processes and data so that they can focus on algorithm development exclusively.

\section{Package architecture}
\label{sec:architecture}

This section describes the main features and the internal architecture of \thename. The complete source code is available at~\cite{opencliper} as a git repository and as a downloadable archive.

\subsection{Features of \thename}
\thename\ is designed as a set of C++ classes which provide three prominent services to the developer: computing device management, data storage and manipulation, and algorithm handling.

\subsubsection{Computing device management}
\paragraph{Straightforward device selection}
One of the most visible drawbacks of OpenCL is that the computing device is not handled automatically, while it is the case with CUDA. OpenCL also introduces the concept of platforms to address the problem of supporting different hardware vendors. It is the developer's job to retrieve and traverse the list of available platforms and devices, and then choose the most appropriate one. With OpenCLIPER, the desired device is selected according to a combination of criteria (e.g. device class, vendor, supported OpenCL version, etc) in a single call.

\paragraph{Widespread computing device support}
OpenCLIPER targets OpenCL version 1.2, which is widely adopted by most significant vendors as of today. Although more recent versions include interesting features, such as shared virtual memory, we plan to continue supporting 1.2 at a minimum.

\subsubsection{Data storage and manipulation}
\paragraph{Automatic use of mapped/pinned host memory}
Data transfers to/from the computing device automatically use mapped host memory (also called pinned memory~\cite{nvidia_best_practices}), so the transfer speed is maximized. Data objects are transferred in a single call.

\paragraph{Support for highly-heterogeneous data}
Other image processing and reconstruction approaches, such as Gadgetron, typically provide a data structure to contain a single N-dimensional array of a given data type (integer, float, etc) the transfer of which, in addition, should be carried out component-wise manually. However, many application fields manage data which does not fit comfortably in that abstraction. Consider, for example, several 3D+t non-uniformly sampled volumes coming from several sensors, each one with various synchronization signals of their own. If this was the case, one would have to create their own \textit{ad-hoc} data structures and handle data transfers component-wise manually as well. OpenCLIPER, however, is agnostic about the internal data organization, so arbitrarily complex data can be transferred to/from the device in a single call. Both OpenCL buffers and images are supported.

\paragraph{Predictable data layout in device memory}	
OpenCLIPER keeps track of data positions and sizes automatically, in both the host and the computing device. A single data set is always aligned and contiguous, even though  it is highly heterogeneous. Data can be processed in batches because the starting position and the size of each component  is known in advance and it is readily available from OpenCL kernels.

\paragraph{Common data formats supported out-of-the-box}
OpenCLIPER supports many image formats as well as volumes in Matlab's \texttt{.mat} format. For unsupported data formats, new readers and writers may be added by deriving from the appropriate class. In addition, OpenCLIPER supports volumes in raw data format as well.

\subsubsection{Algorithm handling}
\paragraph{Automatic kernel loading and compiling}
In OpenCLIPER, kernel loading and compiling, as well as error checking and reporting,  is automatically done in a single call, even for kernels scattered among multiple source files. It also keeps track of kernels at run time so the developer has them readily available by name.

\paragraph{Standard interface to kernel launching}
OpenCLIPER is designed to ease the job of launching kernels;  to this end, the concept of process is introduced. Processes have a customizable but standard calling interface, so that any kernel is treated by means of the three following actions:  set its input and output data sets, set its parameters, and launch. Since it is quite common that kernels need some kind of initialization before doing the actual job, and this initialization may be costly, OpenCLIPER separates kernel initialization from kernel launching, so that performance is not compromised. Processes can be chained at no cost (setting outputs from a stage as inputs for the next one is zero-copy).

\paragraph{Single source code}
Being OpenCL-based, the same source code is valid for every computing device (as long as there is an OpenCL implementation for it), as opposed to CUDA-based approaches.

\subsection{Description of \thename}
\label{subsec:description}

\begin{figure*}
    \begin{center}
	\begin{scriptsize}
	    \includegraphics[width=\textwidth]{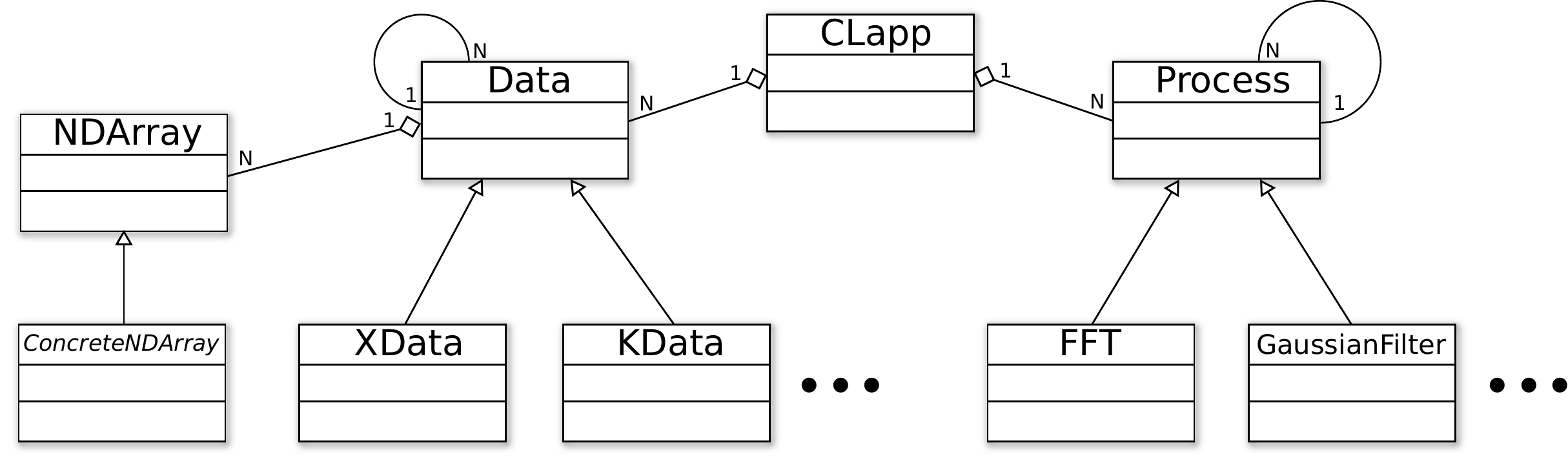}
	\end{scriptsize}
    \end{center}
    \caption{Simplified class diagram of \thename. Italics denote a templated class; dots denote ``as many other classes as needed''; \textls[-70]{\raisebox{.25pt}{\textemdash}$\triangleright$} denotes inheritance; \textls[-80]{\raisebox{.25pt}{\textemdash}$\diamond$} denotes aggregation; numbers denote multiplicity.}
    \label{fig:class_diagram}
\end{figure*}

Figure~\ref{fig:class_diagram} depicts a \textit{simplified} class diagram where the main classes involved can be seen at a glimpse. These classes are described in section~\ref{subsec:description}, and their usage is discussed in section~\ref{subsec:usage}.

\begin{itemize}
    \item \texttt{CLapp} is the main class of \thename. It acts as an interface to the OpenCL device and, as such, stores information about the current OpenCL platforms and devices, their associated command queues and so on. It also contains the list of data objects to be processed in the computing device. 
    \texttt{CLapp} also deals with memory management of the computing device, as well as with data transfers to/from it. 

    \item \texttt{Data} is an abstract class. Objects of its derived classes contain a set of images, volumes or $n$-dimensional data in the most general case. Each (derived from) \texttt{Data} object consists of one or more \texttt{NDArray} objects, which need not be equal in sizes or dimensions. This way, a single acquisition containing heterogeneous data may be stored in a single object. 

    \texttt{Data} is meant to be specialized for the problem to be solved, so final implementations may contain as many as necessary data types, be them real (such as for echo or radiograph data) or complex (such as for $K$-space data from an MRI scanner). \thename\ provides two general-purpose specializations out-of-the-box called \texttt{XData} (for data with a direct physical interpretation) and \texttt{KData} (for $K$-space data). Readers and writers are provided for the most commonly-used image formats (via the DevIL library~\cite{devil}), as well as for Matlab's \texttt{.mat} format.

    \item \texttt{NDArray} represents a signal, image, volume or any $n$-dimensional data structure. It is defined as an abstract class because it does not know the specific (machine) data type in which data are stored. Thus, it is limited to common attributes and methods for all possible data types, and to creation of objects of the actual container class, which is \texttt{ConcreteNDArray}. This way, implementors of new \texttt{Data} specializations do not have to deal with the deep details of data storage and class coupling is kept to a minimum.

    \item \texttt{ConcreteNDArray} is a templated class which stores the actual data. It is not meant to be used by \thename\ users but just to contain raw data and details which depend on the machine data type.

    \item \texttt{Process} is an interface to algorithms which process data. As such, it is an abstract class which developers should derive from to implement their own processes. Its purpose is to provide a standard front-end to algorithms, so that no prior knowledge about their internals is needed to start working with them. 
\end{itemize}

Overall, \thename\ is designed to simplify the handling of computing device, data and algorithms while still allowing fast processing of data. To this end, pointers to raw data are always available, both in host memory and in device memory, and process launching is designed to be as immediate as possible.

While having three classes involved in data storage (\texttt{Data}, \texttt{NDArray} and \texttt{ConcreteNDArray}) may seem redundant, it has a number of advantages: first, data sets containing information from different sources may be easily built; second, low-level particularities related to machine data representation are isolated from the user (e.g. user classes do not need to be templated), thus enhancing class decoupling and, third, support for new data types may be more easily added without affecting the rest of the code.

\subsection{Usage of \thename}
\label{subsec:usage}

A typical usage of \thename\ follows these simple steps (see listing~\ref{listing:simple_main}):

\begin{enumerate}
\setcounter{enumi}{-1}
    \item Create a \texttt{CLapp} object.
    \item Initialize the computing device. The user may impose a specific platform, device or device class, according to several criteria, or leave the choice to the framework.
    \item Load and compile OpenCL kernels. This is done in a single call; the framework takes care of creating program objects, indexing kernels, compiling and linking for the chosen device and making them available to the user. If compilation fails, the error log is automatically at user disposal.
    \item Create an input \texttt{Data} object. Data may be loaded from files (JPEG, TIFF, PNG, and other usual image formats are supported, as well as Matlab's \texttt{.mat} format) or just be allocated empty in memory.
    \item Create an output \texttt{Data} object.
    \item Register input and output in the \texttt{CLapp} object. This also sends the data to the computing device.
    \item Create process object(s) and set their input/output to some of the previously registered \texttt{Data} objects.
    \item Initialize and launch processes. As previously discussed, initialization is separated from process launching. Parameters in each call may be varied.
    \item Get data  from the computing device back to the host device. 
    \item Save processed data. As when loading input data, commonly used image formats and Matlab's \texttt{.mat} files are supported.
    \item Clean up allocated resources.
\end{enumerate}

Listings~\ref{listing:simple_main}, \ref{listing:simple_process_header}, \ref{listing:simple_process} and~\ref{listing:simple_kernel} show a simple intensity inverting filter built with \thename\ (\texttt{using namespace} directives are used for brevity). It can be seen that most of the described steps translate in one or two lines of code in listing~\ref{listing:simple_main}, and that C++ exceptions are used to ease error handling. 

Listings~\ref{listing:simple_process_header} and~\ref{listing:simple_process} define the process class while listing~\ref{listing:simple_kernel} describes the actual computation to be executed on each image pixel, in this case, an inversion of contrast.  As can be observed, \thename's main goal is to allow developers to focus on their core work, i.e.  kernel development, while housekeeping is handled by the framework in a fast and simple manner.

\begin{scriptsize}
\begin{lstlisting}[language=C++,linewidth=\columnwidth,breaklines=true,frame=single,label=listing:simple_main,caption={A simple \thename\ example (main program).}]
#include <OpenCLIPER/XData.hpp>
#include <OpenCLIPER/processes/examples/Negate.hpp>
#include <iostream>
#include <string>

using namespace OpenCLIPER;
int main(int argc, char *argv[]) {
  // Step 0: get a new OpenCLIPER app
  std::shared_ptr<CLapp> pCLapp = std::make_shared<CLapp>();

  try {
    // Step 1: initialize computing device
    CLapp::PlatformTraits platformTraits;
    CLapp::DeviceTraits deviceTraits;
    pCLapp->init(platformTraits,deviceTraits);

    // Step 2: load OpenCL kernel(s)
    pCLapp->loadKernels("examples/negate.cl");

    // Step 3: load input data
    std::shared_ptr<Data> pIn(new XData(std::string("Cameraman.tif"), type_index(typeid(realType))));

    // Step 4: create output with same size as input
    std::shared_ptr<Data> pOut(new XData((dynamic_pointer_cast<XData>(pIn)), false));

    // Set 5: register input and output in our CL app
    DataHandle inHandle = pCLapp->addData(pIn);
    DataHandle outHandle = pCLapp->addData(pOut);

    // Step 6: create new process bound to our CL app
    // and set its input/output data sets
    std::unique_ptr<Process> pProcess(new Negate(pCLapp));
    pProcess->setInHandle(inHandle);
    pProcess->setOutHandle(outHandle);

    // Step 7: initialize & launch process
    pProcess->init();
    pProcess->launch();

    // Step 8: get data back from computing device
    pCLapp->device2Host(outHandle, SyncSource::BUFFER_ONLY);

    // Step 9: save output data
    auto outputData=dynamic_pointer_cast<XData>(pCLapp->getData(outHandle));
    outputData->save("output.png", SyncSource::BUFFER_ONLY);

    // Step 10: clean up
    pProcess.reset(nullptr);
    pCLapp->delData(inHandle);
    pCLapp->delData(outHandle);
    pCLapp = nullptr;
  } catch (std::exception& e) {
        std::cerr << "Error: " << e.what() << std::endl;
  }
}
\end{lstlisting}
\begin{lstlisting}[language=C++,linewidth=\columnwidth,breaklines=true,frame=single,label=listing:simple_process_header,caption={A simple \thename\ example (\texttt{Process} class header).}]
#ifndef INCLUDE_OPENCLIPER_NEGATE_HPP_
#define INCLUDE_OPENCLIPER_NEGATE_HPP_

#include <OpenCLIPER/CLapp.hpp>
#include <OpenCLIPER/Process.hpp>

namespace OpenCLIPER {
class Negate : public OpenCLIPER::Process {
  public:
    Negate(): Process() {};
    Negate(std::shared_ptr<OpenCLIPER::CLapp> pCLapp): Process(pCLapp) {};
    void init();
    void launch(ProfileParameters& profileParameters);
};
} //namespace OpenCLIPER
#endif /* INCLUDE_OPENCLIPER_NEGATE_HPP_ */
\end{lstlisting}
\begin{lstlisting}[language=C++,linewidth=\columnwidth,breaklines=true,frame=single,label=listing:simple_process,caption={A simple \thename\ example (\texttt{Process} class implementation).}]
#include <OpenCLIPER/processes/examples/Negate.hpp>

namespace OpenCLIPER {
void Negate::init() {
    kernel = getApp()->getKernel("negate_kernel");
    queue = getApp()->getCommandQueue();
}

void Negate::launch(ProfileParameters profileParameters) {
    // Set input and output OpenCL buffers on device memory
    cl::Buffer* pInBuf = getInput()->getContiguousMemoryDeviceBuffer();
    cl::Buffer* pOutBuf = getOutput()->getContiguousMemoryDeviceBuffer();
    
    // Set kernel parameters
    kernel.setArg(0, *pInBuf);
    kernel.setArg(1, *pOutBuf);

    // Set kernel work items size: number of pixels to process is image width x height
    cl::NDRange globalSizes = {NDARRAYWIDTH(getInput()->getNDArray(0)) * NDARRAYHEIGHT(getInput()->getNDArray(0))};

    // Execute kernel
    queue.enqueueNDRangeKernel(kernel, cl::NullRange, globalSizes, cl::NDRange(), NULL, NULL);
}
}
\end{lstlisting}
\begin{lstlisting}[language=C++,linewidth=\columnwidth,breaklines=true,frame=single,label=listing:simple_kernel,caption={A simple \thename\ example (kernel).}]
#include <OpenCLIPER/kernels/clKernelDefs.h>

__kernel void negate_kernel(__global realType* input, __global realType* output) {
    int num = get_global_id(0);
    output[num] = (1.0 - input[num]);
}
\end{lstlisting}
\end{scriptsize}

\section{Case study: MRI reconstruction}
\label{sec:case_study}

In what follows, all the calculations have been carried out on a Intel(R) Corporation Core(TM) i7-4790 CPU @ 3.60GHz with 16 GB of RAM), running  Ubuntu 16.04 LTS, host compiler gcc 5.4, OpenCL version 1.2. The GPU used is a GeForce GTX 970, CUDA 7.5 and CUDA compiler nvcc 7.5.17.

\subsection{Sensitivity-weighted multicoil image reconstruction}

MRI reconstruction is a prominent application of medical imaging nowadays. Briefly outlined, data coming from the MRI scanner belongs in the so-called $K$-space, which is the Fourier-transformed space of the real image space (the $X$-space from now on). In addition, current scanners have several antennas and may carry out parallel acquisition and reconstruction, so a single acquisition consists of $N$ $K$-space images $Y_i$, one per antenna. Each antenna $i$ is most sensitive to a particular part of its surrounding space, which is defined by a sensitivity map $S_i$. Therefore, the process of reconstructing an $X$-space image $M$ from the scanner data can be described as:

\begin{equation}
    M=\sum_{i=1}^N {S_i^* \mathcal{F}^{-1}\left( Y_i \right)}
    \label{eq:reconstruction}
\end{equation}

\noindent where $S_i^*$ stands for the complex conjugate of $S_i$.

Translating expression~\ref{eq:reconstruction} into a C++ program using \thename\ is very similar in essence to the example shown in section~\ref{subsec:usage} (notice that  listing~\ref{listing:simple_main} in that section is the equivalent to  listing~\ref{listing:recon_main} in this section and the same goes for listings ~\ref{listing:simple_process} and ~\ref{listing:recon_process}). Now, complementarily, we show some other capabilities of \thename:

\begin{itemize}
    \item \textit{Platform and device traits}: selection of the computing device may be left up to \thename\ as in listing~\ref{listing:simple_main}, or be specified by the user through various criteria such as OpenCL version, device type, platform/device vendor, etc.
    \item \textit{Reading/writing files from Matlab}: Data objects may be constructed right away from Matlab's \texttt{.mat} files and, accordingly, saved to as well.
    \item \textit{Multiple CL program loading}: CL kernels may reside in multiple input files. They are automatically compiled and readily available to the user (indexed by name) in a single step.
    \item \textit{Error log from CL compiler}: If kernel compilation fails, the error log is immediately available in the description of the generated exception.
\end{itemize}

For space reasons, the kernels are not shown here (kernels would be allocated in files \texttt{complexElementProd.cl} and \texttt{xImageSum.cl} as indicated in listing~\ref{listing:recon_main}). Details on the actual operations can be found in \cite{opencliper}. Note also that the Fourier transform uses the clFFT library~\cite{clfft} and is included in \thename\ as a process which depends on it.

\begin{scriptsize}
\begin{lstlisting}[language=C++,linewidth=\columnwidth,breaklines=true,frame=single,label=listing:recon_main,caption={MRI reconstruction with \thename\ (main program).}]
#include <OpenCLIPER/OpenCLIPERDataModel.hpp>

using namespace OpenCLIPER;

int main(int argc, char *argv[]) {
  // Get a new OpenCLIPER app
  std::shared_ptr<CLapp> pCLapp = std::make_shared<CLapp>();

  try {
    // Select CPU as the computing device
    CLapp::PlatformTraits platformTraits;
    CLapp::DeviceTraits deviceTraits;
    deviceTraits.type=CLapp::DEVICE_TYPE_CPU;
    pCLapp->init(platformTraits,deviceTraits);

    // Load OpenCL kernel(s)
    std::vector<std::string> kernelFileNames = {"complexElementProd.cl", "xImageSum.cl"};
    pCLapp->loadKernels(kernelFileNames);

    // Load input data from Matlab file
    vector<string> matlabVars = {"KData", "SensitivityMaps"};
    std::shared_ptr<Data> pInputKData(new KData("MRIdata.mat", matlabVars));

    // Create output with suitable size
    std::shared_ptr<Data> pOutputXData(new XData(dynamic_pointer_cast<KData>(pInputKData)));

    // Register input and output in our CL app
    // (data is sent to the computing device automatically)
    DataHandle inHandle = pCLapp->addData(pInputKData);
    DataHandle outHandle = pCLapp->addData(pOutputXData);

    // Create new process, set its input/output data sets 
    // and bind it to our CL app
    std::unique_ptr<Process> pProcess(new SimpleMRIRecon(pCLapp));
    pProcess->setInHandle(inHandle);
    pProcess->setOutHandle(outHandle);

    // Initialize & launch process
    pProcess->init();
    pProcess->launch();

    // Get data back from computing device
    pCLapp->device2Host(outHandle, SyncSource::BUFFER_ONLY);

    // Save output data
    auto outputData=dynamic_pointer_cast<XData>(pCLapp->getData(outHandle));
    outputData->matlabSave("outputFrames.mat", "XData", SyncSource::BUFFER_ONLY);

    // Clean up
    pProcess.reset(nullptr);
    pCLapp->delData(inHandle);
    pCLapp->delData(outHandle);
    pCLapp = nullptr;
  }
  catch (std::exception& e) {
    std::cerr << "Error: " << e.what() << std::endl;
  }
}

\end{lstlisting}
\begin{lstlisting}[language=C++,linewidth=\columnwidth,breaklines=true,frame=single,label=listing:recon_process,caption={MRI reconstruction with \thename\ (\texttt{Process} class).}]
#include <OpenCLIPER/processes/SimpleMRIRecon.hpp>

namespace OpenCLIPER {
void SimpleMRIRecon::init() {
  pProcInvFFT.reset(new FFT(getApp()));
  pProcInvFFT->setInHandle(getInHandle());
  pProcInvFFT->setOutHandle(getInHandle());
  pProcInvFFT->init();

  pProcSensMapProd.reset(new ComplexElementProd(getApp()));
  pProcSensMapProd->init();

  pProcAddXImages.reset(new XImageSum(getApp()));
  pProcAddXImages->init();
}

SimpleMRIRecon::~SimpleMRIRecon() {
  pProcInvFFT.reset(nullptr);
  pProcSensMapProd.reset(nullptr);
  pProcAddXImages.reset(nullptr);
}

void SimpleMRIRecon::launch(ProfileParameters profileParameters) {
  // Step 0: Inverse FFT of initial KData in place 
  auto launchParmsInvFFT = make_shared<FFT::LaunchParameters>(FFT::BACKWARD);
  pProcInvFFT->setLaunchParameters(launchParmsInvFFT);
  pProcInvFFT->launch(profileParameters);

  // Step 1: product of x-images (result of step 0) by sensitivity maps, in place
  DataHandle sensMapsDataHandle = (dynamic_pointer_cast<KData>(getApp()->getData(getInHandle())))->getSensMapsHandle();
  pProcSensMapProd->setInHandle(getInHandle());
  pProcSensMapProd->setOutHandle(getInHandle());

  // Set parameters: handle of registered sensitivity maps and conjugate sensitivity maps before multiply
  auto launchParamsSensMapsProd = make_shared<ComplexElementProd::LaunchParameters>(ComplexElementProd::conjugate, sensMapsDataHandle);
  pProcSensMapProd->setLaunchParameters(launchParamsSensMapsProd);
  pProcSensMapProd->launch(profileParameters);

  // Step 2: addition of all x-images of same frame (number of x-images per frame is the number of coils used for image adquisition)
  pProcAddXImages->setInHandle(getInHandle());
  pProcAddXImages->setOutHandle(getOutHandle());
  
  pProcAddXImages->launch(profileParameters);
}
}
\end{lstlisting}
\end{scriptsize}

In summary, equation~\ref{eq:reconstruction} translates into a process \texttt{SimpleMRIRecon}, which calls subprocesses that do the inverse FFT, multiply the result by the complex conjugate of the sensitivity maps and sum it all to give the final result (there is zero data copying between processes). Therefore, processes can be thought of as mathematical operators. This metaphore is intentional, and allows developers to think of their algorithms as the mathematical expressions they originate from.

Listings~\ref{listing:recon_main} and~\ref{listing:recon_process} also show how a process can (and should) be split in what needs to be done only once (such as initialization work) and what is to be done in each call to the process (the real work). To illustrate this, let us focus on the FFT process (which internally uses the clFFT library). Like many FFT implementations, clFFT needs an initialization step (called plan baking) which takes much longer than the FFT calculation itself. To reflect this, the \texttt{init()} method of the FFT process (which is called once) does the plan baking, while the \texttt{launch()} method simply performs the FFT calculation itself.

\subsection{Performance comparison with other frameworks: Sum of Squares}

Our purpose is to show that the benefits of \thename ~described in this paper are  accompanied by a performance that is comparable to other state of the art frameworks. To this end, we have carried out another type of reconstruction for multicoil images; in this case, we have used the so-called Sum of Squares, i.e., we have added the squared modulus of the image in each coil (referred to hereafter as RSS, from \textit{root sum of squares}). The frameworks used for comparison have been BART \cite{bart}  and Gadgetron \cite{gadgetron}. These frameworks have been downloaded  from their respective websites; then, they have been compiled on the linux server mentioned above with no host compiler optimization (device compiler does optimize to its maximum).

\begin{table}
\begin{center}
\begin{tabular}{|c|c||c|c||c|c|} \hline
\multicolumn{2}{|c||}{BART} & \multicolumn{2}{|c||}{Gadgetron} & \multicolumn{2}{|c|}{\thename} \\ \hline
FFTW & RSS & FFTW & RSS & clFFT & RSS \\ \hline
19.03 & 5.47 & 7.10 & 6.79 &   24.97 & 3.89 \\ \hline
\end{tabular}
\caption{Figures on CPU (in msec).}
\label{tab:PerformanceFrameworksCPU}
\end{center}
\end{table}

\begin{table}
\begin{center}
\begin{tabular}{|c|c||c|c||c|c|} \hline
\multicolumn{2}{|c||}{BART} & \multicolumn{2}{|c||}{Gadgetron} & \multicolumn{2}{|c|}{\thename} \\ \hline
cuFFT & RSS & cuFFT & RSS & clFFT & RSS \\ \hline
0.011 & 0.277 & 0.015 & 1.687 & 1.361 &  0.252 \\ \hline 
\end{tabular}
\caption{Figures on GPU (in msec).}
\label{tab:PerformanceFrameworksGPU}	
\end{center}
\end{table}

The application domain selected for this experiment is cardiac MRI, namely, 2D cine images \cite{Royuela2015a}; the number of images to be reconstructed is 16, with size $160\times160$ and the number of coils is 8. We depart from Cartesian fully sampled raw data in $K$-space. Table \ref{tab:PerformanceFrameworksCPU} shows performance figures of the three frameworks for an average of one hundred reconstruction executions on a CPU; table \ref{tab:PerformanceFrameworksGPU} shows the corresponding results for execution on a GPU (in both, quantities are expressed in miliseconds). As for the FFT, execution times seem substantially worse for \thename\ than for the other frameworks; however, we should keep in mind that, as for this operation, we are actually comparing library performance as opposed to framework performance; specifically, performance results mainly depend of how efficiently the FFT is coded in each external library (these libraries are the highly, assembly-level optimized cuFFT and FFTW for BART and Gadgetron in GPU and CPU, respectively, and OpenCL's equivalent clFFT for \thename, both in GPU and CPU). RSS, on the other hand, is directly coded as kernels in the three alternatives, and it is a sufficiently simple operation so as to render the effect of internal compiler optimization negligible. Therefore, as the tables reveal, ease of use of \thename\ does not bring forth a performance loss but results are pretty much comparable with those obtained from the other two platforms.

\subsection{Performance comparison with multithread options}

\begin{figure}[htb]
	\begin{scriptsize}
	    \includegraphics[width=\columnwidth]{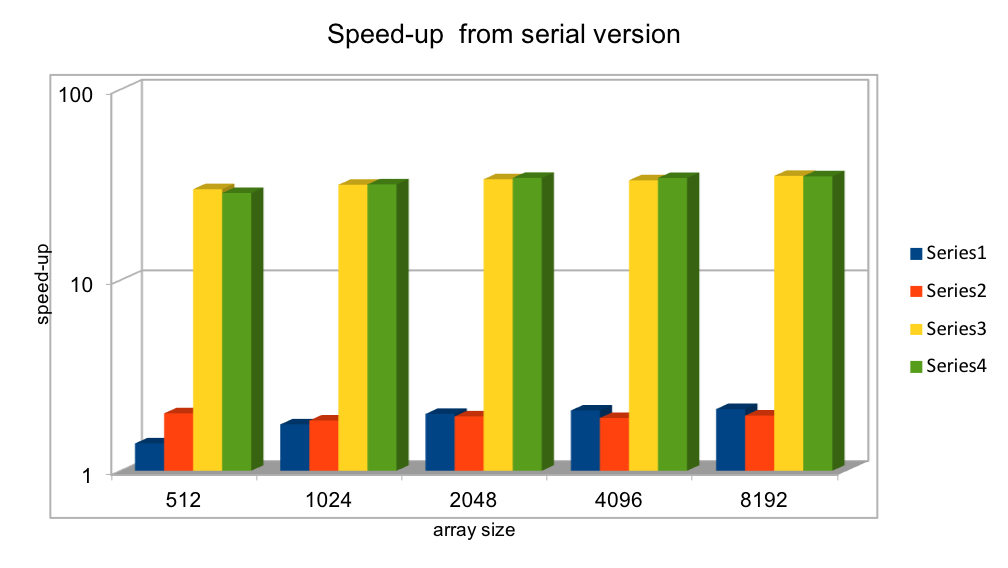}
	\end{scriptsize}
    \caption{Comparison of different devices for a matrix summation operation: Series 1 (blue) OpenMP;   Series 2 (red) \thename\ with CPU as device;  Series 3 (yellow) \thename\ with GPU as device;  Series 4 (green) CUDA; }
    \label{fig:compOMP}
\end{figure}

As a final comparison, we intend to show the benefits of our framework   \thename, with respect to multithread computing. To this end, we have resorted to a simple exercise of adding two matrices and we have measured processing time with respect to the matrices size. We have used as baseline a single thread implementation of the operation. Then four options are compared, namely, the same operation using OpenMP (referred to as Series 1 in figure \ref{fig:compOMP}, blue color), \thename\ using as device CPU (Series 2, red color, in figure \ref{fig:compOMP}), \thename\  using as device GPU (Series 3, yellow color, in the same figure) and CUDA (Series 4, green color); in the four cases the magnitude shown is the speed up with respect to the baseline, i.e., the ratio of the processing time for the baseline and the time for each of the other four schemes (higher is faster). Results are shown, as indicated, in figure \ref{fig:compOMP} for an average of one hundred executions; clearly, multithreading using OpenMP  is not competitive with the GPU alternatives, and it shows comparable results with   \thename\ when CPU is used as the device. With respect to the two GPU implementations, results are very similar, with random variations along the five matrix sizes.

\section{Conclusion and further work}
\label{sec:conclusion}

The main contribution of OpenCL to the programming of dedicated computing devices is to provide a single language and API that covers every device type. Hence, the programs presented here should run unmodified in CPUs, GPUS, DSPs, FPGAs, etc. from any vendor, as long as an OpenCL implementation exists. This is true even when the host program is in binary form since OpenCL kernels are compiled at run-time (when the computing device to use is known). In contrast, CUDA-based programs run exclusively on nVidia GPUs. They cannot even run in the host CPU unless the program includes two versions of the algorithms as, for instance, BART does.

With \thename, every algorithm (process) looks similar to each other by following the 11-point path we have described in section \ref{subsec:usage}. The code separates nicely device administration tasks from algorithmic tasks in an easy-to-trace structured code flow. The programmer should therefore concentrate on the kernel specifics, i.e., on the algorithmic-related issues, while low level tasks are dealt with by the framework. Data structures are defined only once, irrespective of the targeted device; initializations and function execution are also clearly separated for the sake of efficiency. The framework has been satisfactorily tested on CPUs and GPUs from two different vendors. No performance loss is observed when \thename\ is compared with other CUDA-based reconstruction frameworks.

In terms of limitations, at the current stage, \thename\ does not support heterogeneous concurrent computation nor on-the-fly device selection and task partitioning for performance optimization. This is future work we intend to develop.

\section{Acknowledgments}
\label{sec:acknowledgments}
This work is supported by MINECO under grants TEC-2014-57428 and TEC2017-82408-R, as well as by Junta de Castilla y Le\'on under grant VA069U16. 

\bibliographystyle{plain}
\bibliography{bibliography}

\end{document}